\documentclass[twoside,12pt]{article}
\usepackage{epsfig}

\newcommand{\be}{\begin{equation}}
\newcommand{\ee}{\end{equation}}
\newcommand{\bea}{\begin{eqnarray}}
\newcommand{\eea}{\end{eqnarray}}

\begin{document}

\title{ \vspace{1cm} Viscous Effects on Elliptic Flow and Shock Waves}

\author{I. Bouras$^{1}$, L. Cheng $^{1,2}$, A. El$^{1}$, O. Fochler$^{1}$, J. Uphoff$^{1}$, Z. Xu$^{1}$ and C. Greiner$^{1}$
\\
$^1$ Goethe-Universit\"at Frankfurt, Germany \\
$^2$Huazhong Normal University, Wuhan, China
}

\maketitle
\begin{abstract} 
Fast thermalization and a 
strong buildup of elliptic flow of QCD matter 
as found at RHIC are understood as the consequence
of perturbative QCD (pQCD) interactions within the 3+1 dimensional parton
cascade BAMPS. The main contributions stem from pQCD bremsstrahlung 
$2 \leftrightarrow 3 $ processes.
By comparing to Au+Au data of the flow parameter 
$v_2$ as a function of participation number 
the shear viscosity to entropy ratio is dynamically
extracted, which lies in the range of
$0.08$ and $0.2$, depending on the chosen coupling
constant and freeze out condition. Furthermore, first simulations
on the temporal propagation of dissipative shock waves are given.
The cascade can either simulate true ideal shocks as well as initially
diluted, truely viscous shocks, depending on the employed cross sections
or mean free path, respectively. 

\end{abstract}

\section{Introduction: QCD Plasma Thermalization within a \\
pQCD Parton Cascade}
\label{sec:therm}

The values of the elliptic flow parameter $v_2$ measured by the experiments at
the Relativistic Heavy Ion Collider (RHIC) \cite{rhicv2} are (nearly) as large as those obtained from
calculations employing ideal hydrodynamics.
This finding suggests that a fast local equilibration of quarks
and gluons occurs at a very short time scale $\le 1$ fm/c,
and that the locally thermalized state of
matter created, the quark gluon plasma (QGP), behaves as a nearly perfect
fluid. Quarks and gluons
should be rather strongly coupled,
pointing towards a small viscosity to entropy coefficient for the QGP,
which can not be understood by
binary pQCD collisions, because they are too weak.
This has raised the speculation about nonperturbative interactions
as well as about super symmetric representations of Yang-Mills theories
using the AdS/CFT conjecture.
However, in this talk,
we demonstrate that the perturbative QCD (pQCD) can still explain a fast
thermalization of the initially nonthermal gluon system, the large collective
effects of QGP created at RHIC and the smallness of the shear viscosity
to entropy ratio in a consistent manner by using a relativistic pQCD-based
on-shell parton cascade Boltzmann approach of multiparton scatterings
(BAMPS) \cite{XG05,XG07,EXG08,XG08,XGS08,Xu09}. 

\begin{figure}[h!]
\label{dndpt}
\begin{center}
\resizebox{0.45\textwidth}{!}{
  \includegraphics{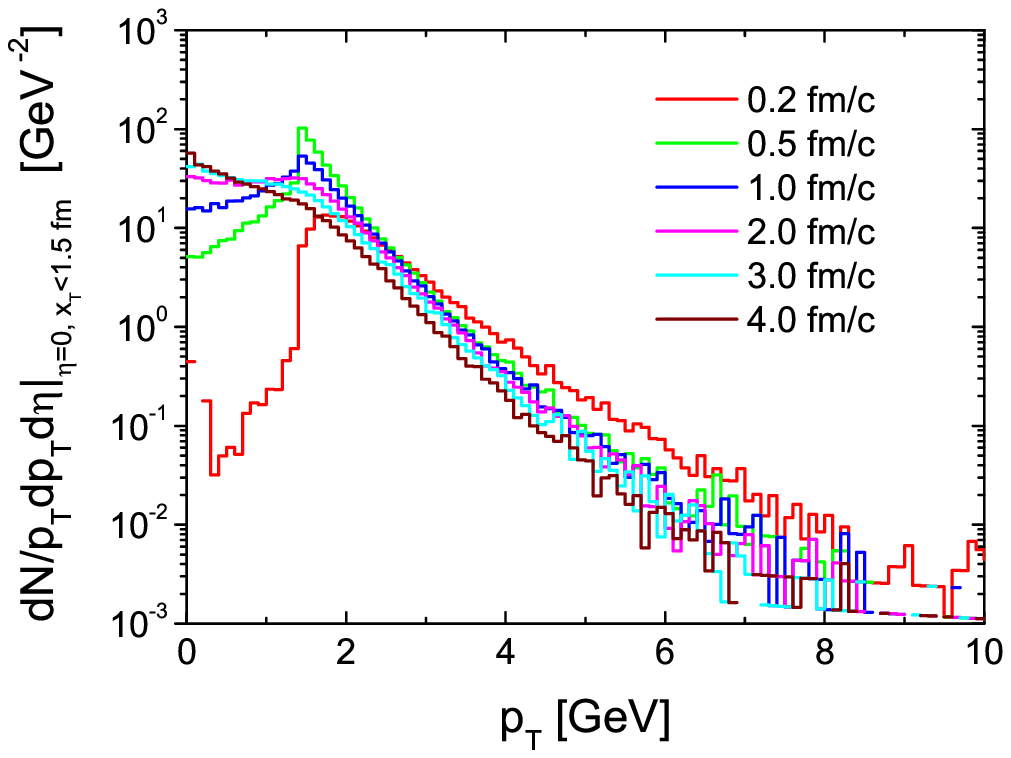}
}
\hspace{\fill}
\resizebox{0.45\textwidth}{!}{
  \includegraphics{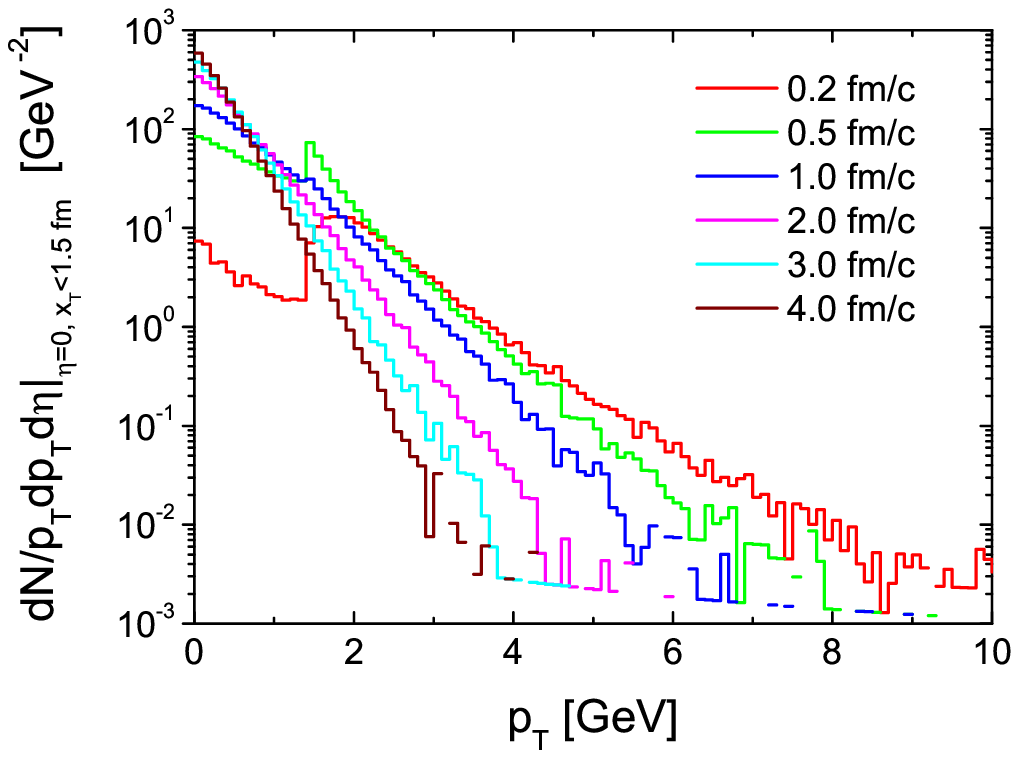}
}
\caption{(Color online) Transverse momentum spectrum in the central region at
different times obtained from the BAMPS simulation
with $gg \to gg$ only (left panel) and including 
$gg \leftrightarrow ggg$ collisions (right panel).
}
\end{center}
\end{figure}

BAMPS is a parton cascade, which solves the Boltzmann transport equation
and can be applied to study, on a semi-classical level, the dynamics of
gluon matter produced in heavy ion collisions at RHIC energies. 
The structure of BAMPS is based on
the stochastic interpretation of the transition rate \cite{XG05},
which ensures full detailed balance for multiple scatterings. BAMPS subdivides
space into small cell units where the operations for transitions are performed.
Gluon interactions included in BAMPS are elastic and screened 
Rutherford-like pQCD $gg\to gg$ 
scatterings as well as pQCD inspired bremsstrahlung
$gg\leftrightarrow ggg$ of Gunion-Bertsch type. 
The matrix elements are discussed in the literature \cite{XG05,XG07,FXG08}.
The suppression of the bremsstrahlung
due to the Landau-Pomeranchuk-Migdal (LPM) effect is taken into account
within the Bethe-Heitler regime employing a step function
in the infrared regime, allowing for independent gluon emissions.

In the default simulations,
the initial gluon distributions are taken in a Glauber geometry 
as an ensemble of minijets with
transverse momenta greater than $1.4$ GeV \cite{XG07}, produced via
semihard nucleon-nucleon collisions.
The later interactions of the gluons are
terminated when the local energy density drops below $1\
\rm{GeV/fm}^3$ \cite{XG07,XGS08} to $0.6\
\rm{GeV/fm}^3$ \cite{Xu09}.
The minijet initial conditions and the subsequent evolution using
the present prescription of BAMPS for two sets of the coupling
$\alpha_s=0.3$ and $0.6$ give nice agreements to the measured 
transverse energy per rapidity over all rapidities
\cite{XGS08,Xu09}.

As a first example the fast thermalization of gluons is demonstrated 
in a local and 
central region which is taken as an expanding cylinder with a radius
of $1.5$ fm and within an interval of space time rapidity
$-0.2 < \eta < 0.2$. Figure 1 shows the varying transverse momentum
spectrum with time obtained from the BAMPS calculations for central Au+Au
collisions at $\sqrt{s}=200$A GeV, with elastic pQCD $gg\to gg$ only
(left panel) and including pQCD-inspired bremsstrahlung
$gg \leftrightarrow ggg$ (right panel), respectively
($\alpha_s=0.3$ is used). One clearly sees that not much happens in the
left panel of figure 1. With only elastic pQCD interactions the gluon
system is initially nonthermal and also stays in a nonthermal state
at late times. The situation is dramatically changed when the inelastic
interactions are included. In the right panel of figure 1 we see that the spectrum reaches an
exponential shape at $1$ fm/c and becomes increasingly steeper at late
times. This is a clear indication for the achievement of local thermal
equilibrium and the onset of hydrodynamical collective expansion with
subsequent cooling by longitudinal work.

The inelastic pQCD based bremsstrahlung and its back reaction are
essential for the achievement of local thermal equilibrium at a short
time scale. The fast thermalization happens in a similar way if color
glass condensate is chosen as the initial conditions.
One of the important messages obtained there is that the hard gluons
thermalize at the same time as the soft ones due to the $ggg\to gg$
process, which is not included in the
so called standard ``Bottom Up'' scenario of thermalization
\cite{EXG08}. The situation is also the same 
when employing PYTHIA based initial conditions \cite{Ch08}.

\section{Shear Viscosity and Elliptic Flow}
\label{sec:shear}

Kinetic equilibration relates to momentum deflection. Large momentum
deflections due to large-angle scatterings will speed up kinetic
equilibration. Whereas the elastic pQCD scatterings favor
small-angle collisions, the collision and emission angles in bremsstrahlung
processes are, for lower invariant, i.e. thermal energies,
rather isotropically distributed due to
the incorporation of the LPM cutoff \cite{XG05,XG07}. 
Hence, although the elastic cross section is still
considerably larger than the inelastic 
one, the given argument is the
intuitive reason why the bremsstrahlung processes are acting
more effectively
in the equilibration of the gluon matter than the elastic interactions.
Quantitatively it was shown in detail in \cite{XG07} that the
contributions of the different processes to momentum isotropization are
quantified by the so called transport rates
$$
R^{\rm tr}_i= \frac{\int \frac{d^3p}{(2\pi)^3} \frac{p_z^2}{E^2} C_i[ f ] -
\langle \frac{p_z^2}{E^2} \rangle \int \frac{d^3p}{(2\pi)^3} C_i [ f ] }{n\,
(\frac{1}{3}- \langle \frac{p_z^2}{E^2} \rangle)} \,,
$$
where $C_i[f]$
is the corresponding collision term describing various interactions,
$i=gg\to gg, gg\to ggg, ggg\to gg$, respectively.
The sum of them gives exactly the inverse of the time scale of momentum
isotropization, which also marks the time scale of overall thermalization.
It turns out that
$R^{\rm tr}_{gg\to ggg}$ is a factor of $3-5$ larger than
$R^{\rm tr}_{gg\to gg}$ over a wide range in the coupling constant, 
which demonstrates the essential role of the
bremsstrahlung in thermal equilibration \cite{XG07}.

\begin{figure}[h!]
\label{v2shv}
\begin{center}
\resizebox{0.45\textwidth}{!}{
  \includegraphics{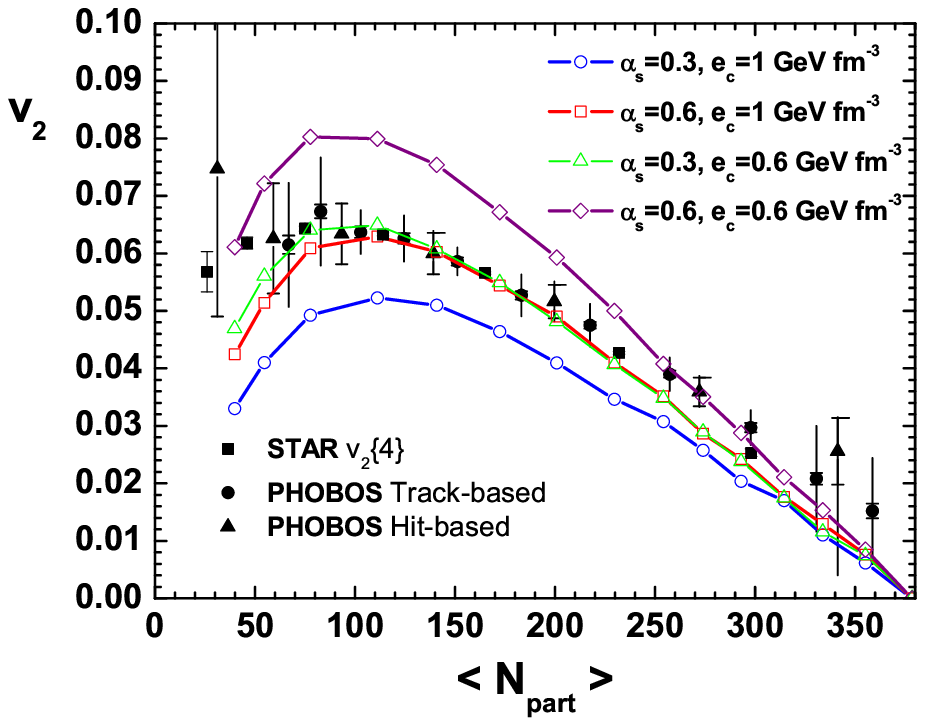}
}
\hspace{\fill}
\resizebox{0.45\textwidth}{!}{
  \includegraphics{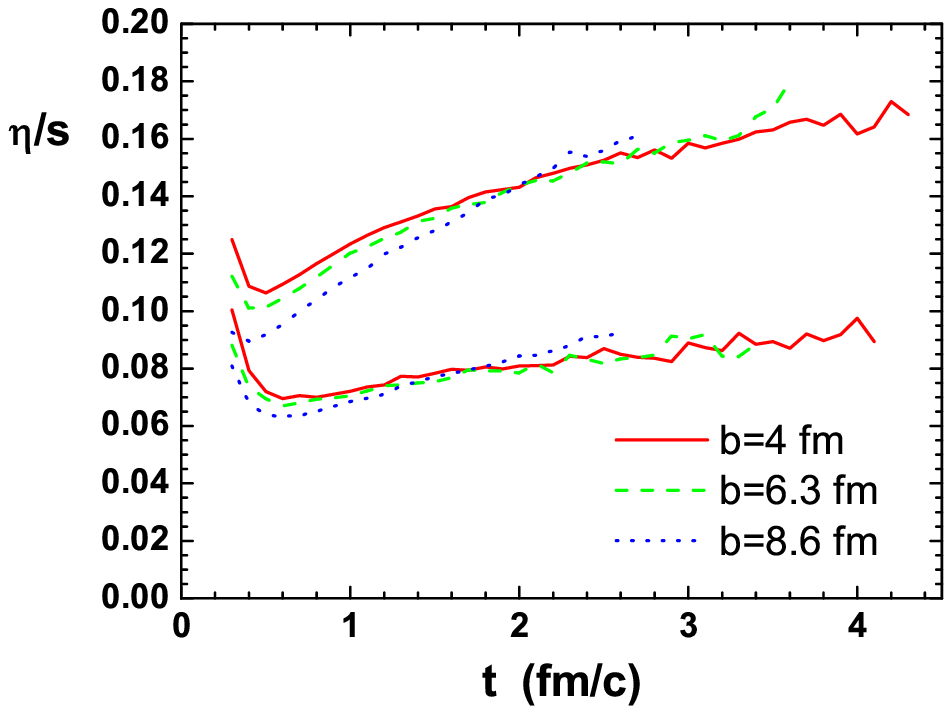}
}
\caption{(color online)Left panel: Elliptic flow vs. $N_{\rm part}$ for Au+Au collisions
at $\sqrt{s_{NN}}=200$ GeV. The points are STAR \cite{star} and PHOBOS 
\cite{phobos} data for charged hadrons within $|\eta| < 0.5$ and $|\eta| < 1$,
respectively, whereas the curves with symbols are results for gluons within 
$|\eta| < 1$, obtained from the BAMPS calculations with 
$\alpha_s=0.3$ and $0.6$ and with two freezeout energy densities,
$e_c=0.6$ and $1$ $\rm{GeV fm}^{-3}$.
Right panel: the shear viscosity to entropy density
ratio $\eta/s$ at the central region during the entire expansion. $\eta/s$
values are extracted from the simulations at impact parameter $b=4$,
$6.3$, and $8.6$ fm. The upper band shows the results with $\alpha_s=0.3$
and the lower band the results with $\alpha_s=0.6$.
}
\end{center}
\end{figure}

Employing the Navier-Stokes approximation the shear
viscosity $\eta$ is directly related to the transport rate \cite{XG08},
$$
\eta \cong \frac{1}{5} n \frac{\langle E(\frac{1}{3}-\frac{p_z^2}{E^2}) \rangle}
{\frac{1}{3}-\langle \frac{p_z^2}{E^2} \rangle} \frac{1}{\sum R^{\rm tr}+ 
\frac{3}{4} \partial_t (\ln \lambda)}\,,
$$
where $\lambda$ denotes the gluon fugacity.
This expression allows to calculate the viscosity dynamically
and locally in a full and microscopical simulation (see Figure 2).
On the other hand,
close to thermal
equilibrium the expression reduces to the more intuitive form 
$
\eta=\frac{4}{15} \, \frac{\epsilon}{\sum R^{\rm tr}}
$
and thus for the shear viscosity to entropy ratio 
$
\frac{\eta}{s}=
\left (5 \beta R^{\rm tr}_{gg \to gg}+ 
\frac{25}{3} \beta R^{\rm tr}_{gg \to ggg} \right )^{-1}
$.
Within the present description bremsstrahlung and its back reaction
lower the shear viscosity to entropy density ratio significantly by
a factor of $7$, compared with the ratio when only elastic collisions
are considered. For $\alpha_s=0.3$ one finds $\eta/s=0.13$
\cite{XG08,EXG08}. To match
the lower bound of $\eta/s=1/4\pi$ from the AdS/CFT
 conjecture $\alpha_s=0.6$ has to be employed.
Even for that case the cross sections are in the order of $1$ mb for a
temperature of $400$ MeV.
The numbers for the viscosity slightly increase up to 20 percent when
calculating it by means of Grads momentum method and using the 2nd order
Israel-Stewart framework of dissipative relativistic hydrodynamics \cite{El08}.
In any case, perturbative QCD interactions
can drive the gluon matter to a strongly coupled system with an
$\eta/s$ ratio as small as the lower bound from the AdS/CFT conjecture.

The elliptic flow $v_2=\langle (p_x^2-p_y^2)/p_T^2 \rangle$ can be
directly calculated from microscopic simulations for Au+Au collisions
at $\sqrt{s}=200$A GeV employing BAMPS. $\alpha_s=0.3$ and $0.6$
and also two values of the critical energy density $e_c$ for the freezeout
are used for comparisons \cite{XGS08,Xu09}. 
For a firm footing we compare our results with
the experimental data, assuming parton-hadron duality. 
The left panel of Figure 2 shows the elliptic flow $v_2$ at midrapidity
for various centralities (impact parameters), compared with the
PHOBOS \cite{phobos} and STAR \cite{star} data.
Except for the central centrality region the results with $\alpha_s=0.6$
agree perfectly with the experimental data, whereas the results with
$\alpha_s=0.3$ are roughly $20\%$ smaller. On the other hand, one also
sees that the $v_2$ results with $\alpha_s=0.3$ and 
$e_c=0.6$ ${\rm GeV fm}^{-3}$ (green curves with open triangles) are 
almost identical with those with $\alpha_s=0.6$ and 
$e_c=1$ ${\rm GeV fm}^{-3}$ (red curves with open squares).
Stronger interactions or longer QGP phase leads to the same final values
of $v_2$. Hence, according to the present study, $\eta/s$ is most probably lying  between $0.2$ and $0.08$.
These findings are in line with a similar study of the Catania group
employing a parton cascade with binary and isotropic interactions 
and large cross
sections \cite{FCTG08}.
In the present case, however, the generation of
the large elliptic flow observed at RHIC is well described by pure
perturbative gluon interactions as incorporated in BAMPS.

We note that adding quark degrees of freedom into 
the dynamical evolution of the QCD matter with a detailed understanding of
the hadronization of gluons and quarks and the
late stage hadronic interactions will further be helpful in explaining 
the viscous facets of the final hadron elliptic flow 
(for a discussion see \cite{Xu09}). Also different picture of initial
condition (eg. color
glass condensate) may also lead to different initially spatial eccentricity,
and, hence, will moderately affect the final value of $v_2$. 
These investigations are underway and will provide more constrains on
extracting $\eta/s$.

\begin{figure}[h!]
\label{shock1}
\begin{center}
\resizebox{0.45\textwidth}{!}{
  \includegraphics{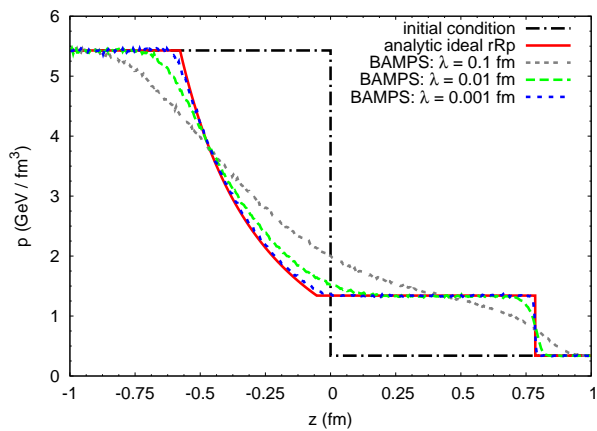}
}
\caption{(Color online) Pressure profile at time $t=1 $ fm/c of a longitudinal 
shock developing from an initial Riemann singularity
with $T_{\mbox{left}}=400$ MeV and $T_{\mbox{right}}=200$ MeV 
for various finite, but increasingly small mean free path of the 
colliding partons. For comparison, the analytical solution of the Riemann
problem for ideal relativistic hydrodynamics is also depicted by the red solid curve.
}
\end{center}
\end{figure}

\section{Dissipative Relativistic Shock Waves}
\label{sec:shocks}

The quenching of gluonic jets can also be self-consistently 
addressed within the same full simulation of BAMPS \cite{FXG08}
including elastic and radiative collisions. 
The nuclear modification factor
$R_{AA}$ of gluons is obtained directly by taking the ratio of the 
final $p_{T}$ spectra to the initial mini-jet spectra.  
A clear suppression of high--$p_{T}$ gluon jets at a roughly constant 
level of $R_{AA}^{\mathrm{gluons}} \approx 0.06$ is found. 
The dominant contribution for the energy loss is the 
bremsstrahlung \cite{FXG08}.
The suppression is approximately a factor of three stronger than the
experimental pion data. However, at present, 
the simulation does not include any light quarks, which are expected
to lose less energy by a significant factor of $4/9$ compared with gluons. 
A more detailed microscopic understanding of the various phenomena
related to high energy jets in the parton matter represents an
intriguing, ongoing project.

\begin{figure}[h!]
\label{shock2}
\begin{center}
\resizebox{0.9\textwidth}{!}{
  \includegraphics{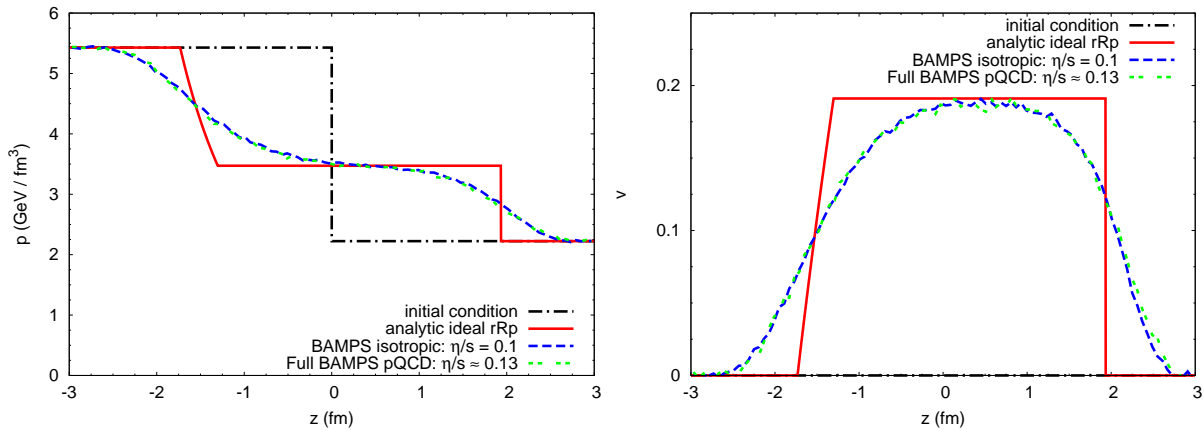}
}
\caption{(Color online) Pressure profile (left panel) 
and velocity profile (right panel) at time $t=3 $fm/c of a longitudinal 
shock developing from an initial Riemann singularity
with $T_{\mbox{left}}=400$ MeV and $T_{\mbox{right}}=320$ MeV 
for a plasma with a fixed shear viscosity to entropy ratio of 0.1 (with binary and isotropic
scattering) and for a plasma with the full pQCD BAMBS simulation ($\alpha_s=0.3$) including bremsstrahlung processes.
The analytical solution of the Riemann
problem for ideal relativistic hydrodynamics is also depicted by the red solid curve.
}
\end{center}
\end{figure}

Experimentally a significant and exciting structure in the two-particle and
three-particle correlations of associated particles of a high energy
jet has been observed, which might be the result of the conical
emission of
propagating Mach cones created by a jet crossing the expanding
medium \cite{Ul07}. Nuclear shock waves have been proposed
since a long time in the context of heavy ion collisions. Recent
calculations invoking ideal hydrodynamics for the QCD matter support 
a picture of creation of Mach
cones associated with a jet loosing significant energy \cite{Be08}.
At present it is yet not possible to see developing Mach cone-like
structures within the  cascade BAMPS, as jet-quenching has to be
further understood and such simulations (will) need intensive computing
time.

An important question to be answered aside is whether relativistic
shock waves can be observed in parton cascade simulations and how
finite (shear) viscosity will alter such a picture.
In the following we report on a very recent study \cite{Bo08}.

\begin{figure}[h!]
\label{shock3}
\begin{center}
\resizebox{0.7\textwidth}{!}{
  \includegraphics{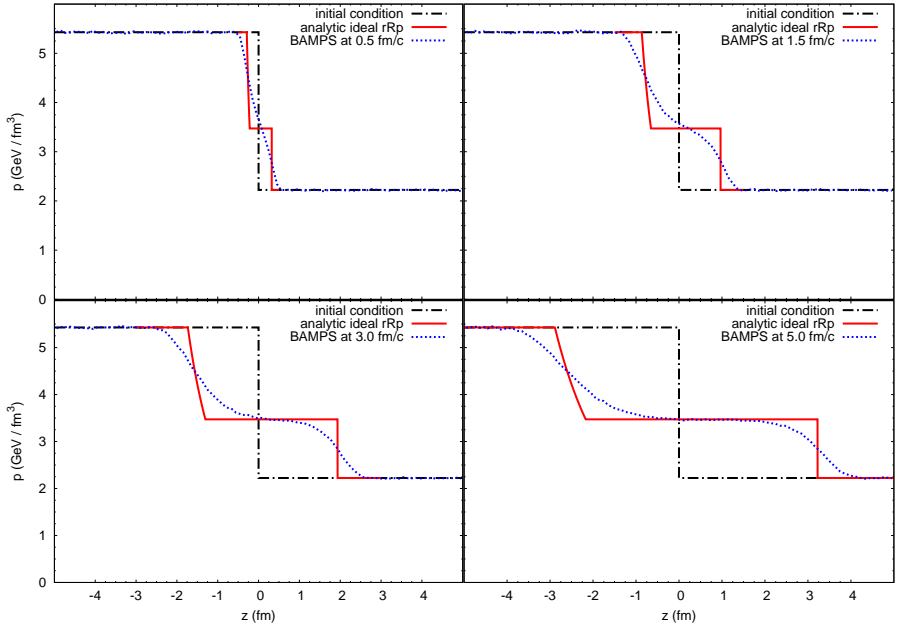}
}

\caption{(Color online) Pressure profile 
at various times ($t=0.5,1.5,3,5$ fm/c) of a longitudinal 
shock 
(with $T_{\mbox{left}}=400$ MeV and $T_{\mbox{right}}=320$ MeV) 
for a plasma with fixed shear viscosity to entropy ratio of 
$\frac{\eta }{s} = \frac{1}{4 \pi }$ refering to the lower bound
of the AdS/CFT conjecture. 
}
\end{center}
\end{figure}

A particular case of a to be created shock wave represents the
well established (relativistic) Riemann problem:
At a particular layer (eg the
x-y-plane) there is an initial discontinuity in pressure and energy density
of thermalized matter. Within an ideal hydrodynamic framework
the further temporal evolution can be solved (semi-)analytically
for a given equation of state. Typically an immediate shock wave
evolves to the side of lower density whereas a rarefaction wave
evolves
to the side of higher density. 

To make the computing time consumption tolerable, for most of the calculations
only $2 \leftrightarrow 2$ collisions with a isotropic cross section are 
employed. Either a constant mean free path in the computational frame via
$$
\lambda_{\mbox{mfp}} \, = \, 
\frac{\gamma(z,t)}{n(z,t)\,  \sigma (z,t)}
$$
or a constant shear viscosity to entropy ratio via
$$
\frac{\eta }{s} \, = \, \frac{16}{45} \,  \frac{\epsilon(z,t)}{n(z,t) \,  
\sigma (z,t) } \,  \frac{1}{4 n (z,t) - n (z,t)
 \ln (\lambda (z,t)) }
$$
can be employed, where $n$ counts the local gluon density,
$\sigma $ the local cross section, $\gamma=1/\sqrt{1-v^2}$ the 
Lorentz factor, $\epsilon $ the local energy density, and
$\lambda$ the local gluon fugacity.

In Figure 3 calculations of a particular example of the Riemann
problem are shown, when a (variety of fixed) mean free path 
(for isotropic binary
collisions) is chosen. Details of the simulations are in ref. \cite{Bo08}.
Shown is the pressure profile at a time $t=1$ fm/c of a longitudinal 
shock developing from an initial Riemann singularity at $z=0$ fm
with $T_{\mbox{left}}=400$ MeV and $T_{\mbox{right}}=200$ MeV. 
The mean free pathes of the 
colliding partons are increasingly small.
The matter would have an unrealistic small shear viscosity to entropy
ratio $\eta / s$ of about $0.04, 0.004, 0.0004$. For comparison, 
the analytical solution of the Riemann
problem for ideal relativistic hydrodynamics is depicted by the red
solid curve. One nicely recognizes how the correct limit
of ideal hydrodynamics is achieved, which is nontrivial! 
Only for tiny shear viscosities the shock waves do resemble that of
the ideal case. The present calculations do present a cornerstone 
for investigating dissipative shock waves within transport simulations.

Figure 4 shows the pressure profile (left panel) 
and velocity profile (right panel) at a time $t=3 $fm/c of a longitudinal 
shock developing from an initial Riemann singularity at $z=0$ fm
with $T_{\mbox{left}}=400$ MeV and $T_{\mbox{right}}=320$ MeV 
for a plasma with a fixed shear viscosity to entropy ratio of 0.1 
(with binary and isotropic scattering) and for a gluon plasma with 
the {\em full} pQCD BAMBS simulation ($\alpha_s=0.3$) including 
bremsstrahlung processes.
Both calculations basically agree as $\eta /s$ in the full simulation
is about 0.13. Hence the full pQCD simulations would give rather
dissipative shock waves. Yet such calculations are, in principle, 
though extermely time consuming, possible. It remains to be seen 
whether Mach cone like structures can be built up and maintained in
realistic 3+1 dimensional calculation when a jet is produced and loses 
energy to the medium. This remains a formidable task for the future.

In Figure 5 the temporal evolution of the
pressure profile of a longitudinal shock 
(with $T_{\mbox{left}}=400$ MeV and $T_{\mbox{right}}=320$ MeV) 
for a plasma with fixed shear viscosity to entropy ratio of the 
celebrated AdS/CFT conjectured value of
$\frac{\eta }{s} = \frac{1}{4 \pi }$ is shown.
For small times $t\leq 2 $ fm/c the situation rather looks like
a diffusion process \cite{Bo08} and one might not really speak about
a (fully) developed shock wave. On the other hand, for longer
times the characteristic shock plateau develops and the overall situation
resembles that of an ideal shock
with selfsimilar properties. A special Knudsen number may 
characterize whether a shock has developed or not \cite{Bo08}. 

In summary, dissipative shocks can be nicely simulated and investigated
with parton cascade BAMPS. Whether Mach cone-like behaviour can be
observed within realistic simulations for RHIC is an exciting issue, 
as the QGP life time is not long and the medium is indeed viscous. 
In addition, cascade simulations
provide an execellent basis for testing numerical algorithms
for the various developing dissipative hydrodynamical theories.

{\bf Acknowledgments:} 
We are grateful to the Center for the Scientific Computing (CSC) at
Frankfurt for the computing resources. This work was supported by
BMBF, DAAD, DFG, GSI and by the 
Helmholtz International Center
for FAIR within the framework of the LOEWE program (Landes-Offensive zur
Entwicklung Wissenschaftlich-\"okonomischer Exzellenz) launched
by the State of Hesse.

\end{document}